\documentclass[%
 reprint,
nofootinbib,
 amsmath,amssymb,
 aps,
prstab,
]{revtex4-2}

\usepackage{graphicx}
\usepackage{dcolumn}
\usepackage{bm}

\usepackage[caption=false]{subfig}

\begin{document}

\preprint{APS/123-QED}

\title{Analysis of a hadron beam in five-dimensional phase space}

\author{A. Hoover}%
 \email{hooveram@ornl.gov}
\author{K. Ruisard}%
 \email{ruisardkj@ornl.gov}
\author{A. Aleksandrov}%
\author{A. Zhukov}%
\author{S. Cousineau}
\affiliation{Oak Ridge National Laboratory, Oak Ridge, Tennessee 37830, USA}

\date{\today}

\begin{abstract}

We conduct a detailed measurement and analysis of a hadron beam in five-dimensional phase space at the Spallation Neutron Source Beam Test Facility. The measurement's resolution and dynamic range are sufficient to image sharp, high-dimensional features in low-density regions of phase space. To facilitate the complex task of feature identification in the five-dimensional phase space, we develop several analysis and visualization techniques, including non-planar slicing. We use these techniques to examine the transverse dependence of longitudinal hollowing and longitudinal dependence of transverse hollowing in the distribution. This analysis strengthens the claim that low-dimensional projections do not adequately characterize high-dimensional phase space distributions in low-energy hadron accelerators.

\end{abstract}

\maketitle

\section{Introduction}\label{sec:introduction}

The beam intensity in hadron linear accelerators is limited by space-charge-driven halo formation --- the emergence of a low-density region of phase space far from a dense core \cite{Gluckstern1994, Wangler1998, Batygin2021} --- and consequent uncontrolled beam loss \cite{Cousineau2015}. In megawatt-class accelerators, the halo density (in two-dimensional phase space) is typically four to six orders of magnitude below the peak density \cite{Aleksandrov2020}. No simulation has reproduced measurements at this level of detail. Although the relevant physics is assumed to be modeled correctly, there remain significant uncertainties in the simulation inputs --- the electromagnetic fields throughout the accelerator and the initial distribution of particles in six-dimensional phase space \cite{Allen2002, Qiang2002, Groening2008}. 

We denote the phase space distribution by $f(x, x', y, y', \phi, w)$; $x$ and $y$ are the transverse positions, $x' = dx / ds$ and $y' = dy / ds$ are the transverse slopes, $s$ is the position along the reference trajectory, $\phi$ is the deviation from the longitudinal position of the synchronous particle (in units of RF degrees), and $w$ is the deviation from the kinetic energy of the synchronous particle. The distribution is typically reconstructed from the set of measured two-dimensional projections $\left\{{f(x, x'), f(y, y'), f(\phi, w)}\right\}$, where each projection is obtained by integrating over the unlisted coordinates:
\begin{equation}
    \begin{aligned}
        f(x, x') &= \iiiint
        f(x, x', y, y', \phi, w) {dy}{dy'}{d\phi}{dw}, \\
        f(y, y') &= \iiiint
        f(x, x', y, y', \phi, w) {dx}{dx'}{d\phi}{dw}, \\
        f(\phi, w) &= \iiiint
        f(x, x', y, y', \phi, w) {dx}{dx'}{dy}{dy'}.
    \end{aligned}
\end{equation}
Given only this information, the reconstruction must take the following maximum-entropy form \cite{Wong2022-tomography}:
\begin{equation}\label{eq:reconstruction}
    f(x, x', y, y', \phi, w) = f(x, x') f(y, y') f(\phi, w). 
\end{equation}

Direct high-dimensional measurements have been demonstrated, albeit at low resolution and dynamic range \cite{Cathey2018}. The most immediate and straightforward use of such measurements is as a seed for macro-particle simulations. In this case, no analysis of the initial distribution is required. Alternatively, these measurements may be analyzed to identify features in high-dimensional phase space --- features invisible to typical diagnostics. This task is critical to fully understanding the limitations of Eq.~\eqref{eq:reconstruction} when predicting subsequent beam evolution. Additionally, explaining the origin of high-dimensional features may elucidate the dynamics upstream of the measurement plane. This is the path taken in the present study, which builds upon the following work.

The first six-dimensional phase space measurement characterized a 2.5 MeV H$^-$ ion beam generated by a radio-frequency quadrupole (RFQ) at the Spallation Neutron Source (SNS) Beam Test Facility (BTF) using four transverse slits, a dipole-slit energy spectrometer, and a bunch shape monitor (BSM) \cite{Cathey2018}. The resolution ($\approx 11$ points per dimension) and dynamic range ($\approx 10^1$) were relatively low, even with 32 hours of measurement time; therefore, as part of a preliminary investigation, lower-dimensional scans were used to examine smaller regions of phase space. Masking the beam in the transverse plane before measuring the energy distribution --- measuring $f(w \mid x{=}x'{=}y{=}y'{=}0)$ --- revealed a bimodal energy distribution near the transverse core. Importantly, this feature was not visible in the full projection $f(w)$, which was unimodal. The correlation's five-dimensional nature was briefly explored by varying the number of slits inserted into the beam and by varying the location of a single slit with the others held fixed; both led to pronounced changes in the energy distribution. Repeating the measurement at different beam intensities demonstrated that space charge drives this dependence.

The transverse-longitudinal correlations observed in \cite{Cathey2018} were subsequently studied in \cite{Ruisard2020}. The dependence of the longitudinal phase space on $x$ and $x'$ was mapped by measuring $f(x, x', \phi, w \mid \tilde{y}{=}0)$, where $\tilde{y}$ is the BSM wire position (corresponding approximately to $y'$ at the measurement plane). The measurements were also compared to an RFQ simulation, which predicted a similar dependence of the energy distribution on the transverse coordinates. Following the argument in \cite{Cathey2018} that the longitudinal hollowing develops in the MEBT, particle-in-cell simulations were used in \cite{Ruisard2021} to explore the longitudinal hollowing of a Gaussian beam during free expansion. These simulations illuminated the fact that hollowing is a natural consequence of charge redistribution caused by nonlinear space charge forces. However, in the ``realistic'' beam generated by the RFQ simulation, the correlations were already present at the end of the RFQ and showed little evolution in the MEBT. Therefore, it was concluded that this feature likely develops in the RFQ.

In this paper, we continue to refine our image of the initial phase space distribution in the BTF. In particular, we obtain a nearly complete description of the distribution by measuring $f(x, x', y, y', w)$. This five-dimensional measurement, described in Section~\ref{sec:five-dimensional-phase-space-measurement}, captures all significant inter-plane correlations in the initial beam\footnote{The lack of longitudinal focusing in the BTF results in rapid debunching; a strong linear correlation between the phase $\phi$ and energy $w$ develops before the first measurement station.\label{fn:1}} and provides unprecedented detail: the resolution and dynamic range are sufficient to image sharp, high-dimensional features in low-density regions of phase space. To facilitate the complex task of feature identification in the five-dimensional phase space, we develop several analysis and visualization techniques in Section~\ref{sec:results}, including non-planar slicing. In Section~\ref{sec:results-a}, these techniques are used to re-examine the longitudinal hollowing described above. In Section~\ref{sec:results-b}, we pivot to the transverse phase space and its dependence on the longitudinal coordinates, reporting a transverse hollowing that likely develops in the MEBT and is independent of the longitudinal hollowing in the RFQ. In Section~\ref{sec:discussion}, we discuss the use of five-dimensional measurements in future research.

\section{Five-dimensional phase space measurement}\label{sec:five-dimensional-phase-space-measurement}

A detailed description of the BTF is available in \cite{Zhang2020}. The system consists of an RF-driven H$^-$ ion source, 65 keV low-energy beam transport (LEBT), and 402.5 MHz radio-frequency quadrupole (RFQ), all identical to the components in the SNS. These are followed by a 2.5 MeV medium-energy beam transport (MEBT) which is longer than the SNS design and contains no re-bunching cavities. The lattice ends with a 9.5-cell FODO transport line.

\begin{figure}
    \centering
    \includegraphics[width=\columnwidth]{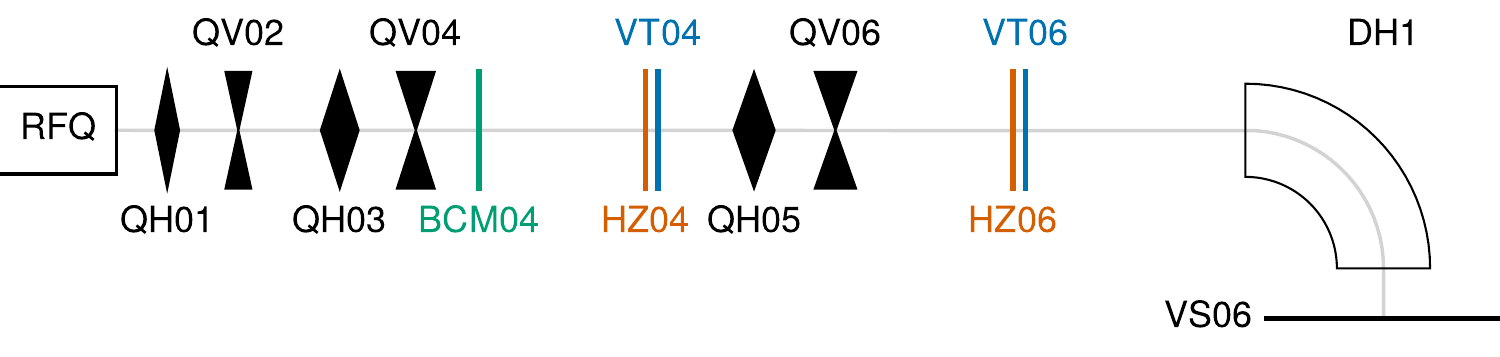}
    \caption{Layout of the first 3.6 meters of the SNS-BTF MEBT, starting from the end of the RFQ. Shown are six quadrupoles (QH01, QV02, QH03, QV04, QH05, QV06), two vertical slits (VT04, VT06), two horizontal slits (HZ04, HZ06), a beam current monitor (BCM04), a 90-degree dipole (DH1), and a view screen (VS06).}
    \label{fig:layout}
\end{figure}

The BTF houses two measurement stations. The first is located 1.3 meters downstream of the RFQ; the second is located after the FODO line. Each station consists of four transverse slits (two horizontal, two vertical) and a 90-degree dipole bend followed by a scintillating screen, as shown in Fig.~\ref{fig:layout}. In this setup, it is possible to measure the five-dimensional distribution $f(x, x', y, y', w)$ using the screen and three upstream slits: one horizontal slit selects $y$; two vertical slits select $x$ and $x'$; $y'$ is a function of $y$ and the vertical position on the screen, $w$ is a function of $x$, $x'$, and the horizontal position on the screen. The transformation from slit-screen coordinates to phase space coordinates is given in Eq.~\eqref{eq:transform}. The measurement is efficient: two dimensions are measured in a single shot. The reduction in the number of scanning slits affords a higher resolution ($>64$ points per dimension) and dynamic range ($>10^3$) than the six-dimensional measurement.

\begin{figure}
    \centering
    \includegraphics[width=\columnwidth]{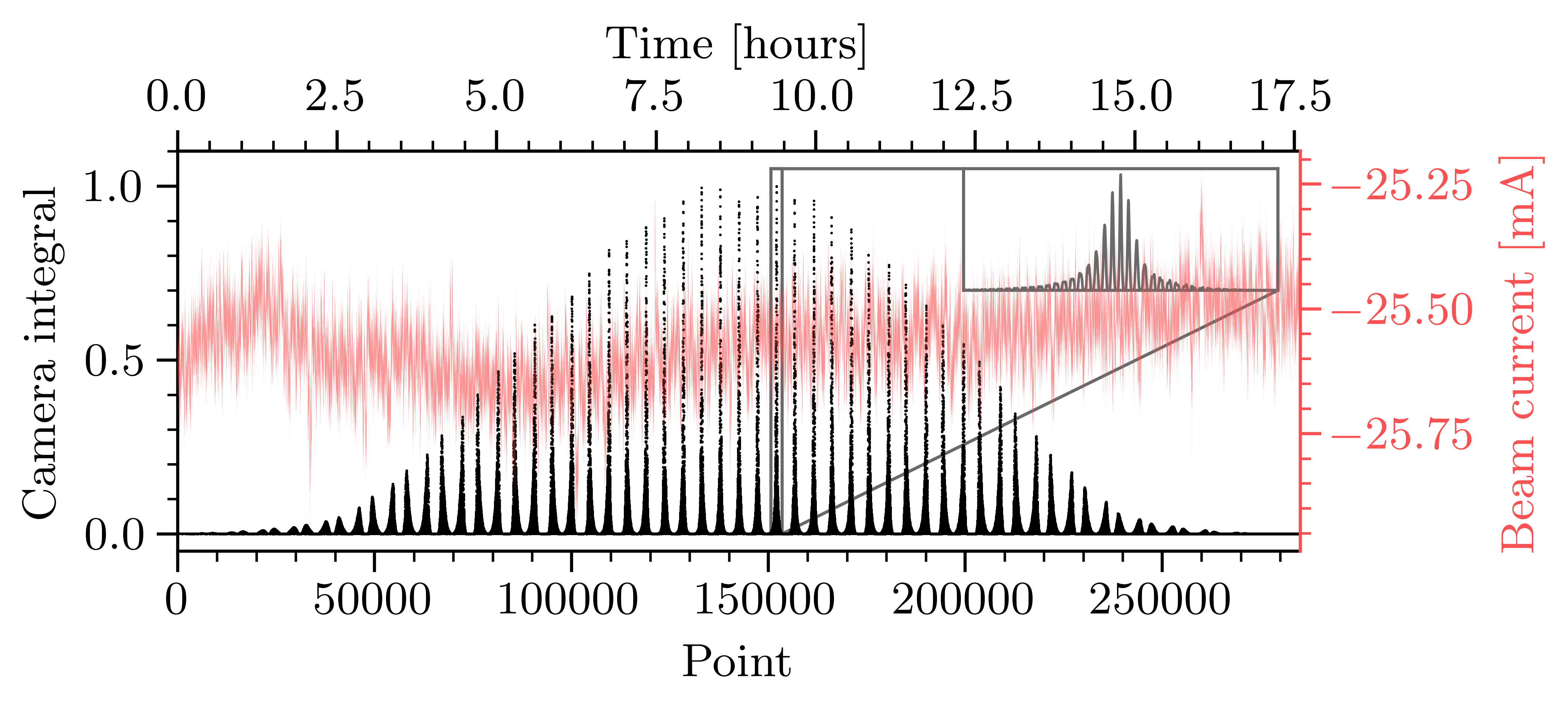}
    \caption{Camera integral (from screen VS06) and beam current (from current monitor BCM04) during the five-dimensional measurement.}
    \label{fig:scan_a}
\end{figure}

We will primarily examine a single measurement in this paper. A rectilinear scan pattern was employed with a linear correlation between $x$ and $x'$ to align with the $x$-$x'$ distribution. The corners of the $x$-$x'$ grid were clipped, leading to a moderate reduction in scan time. The scan was performed as a series of ``sweeps'' in which the vertical slits were held stationary while the horizontal slit was moved continuously across the beam. During each sweep, the screen image was saved on each beam pulse (5 Hz repetition rate) in addition to scalar quantities such as the slit positions and beam current. The camera integral (image brightness) and beam current during the measurement are displayed in Fig.~\ref{fig:scan_a}. The average current during the beam pulse was -25.57 mA, as measured by a beam current monitor (BCM04 in Fig.~\ref{fig:layout}).
\begin{figure}
    \centering
    \includegraphics[width=\columnwidth]{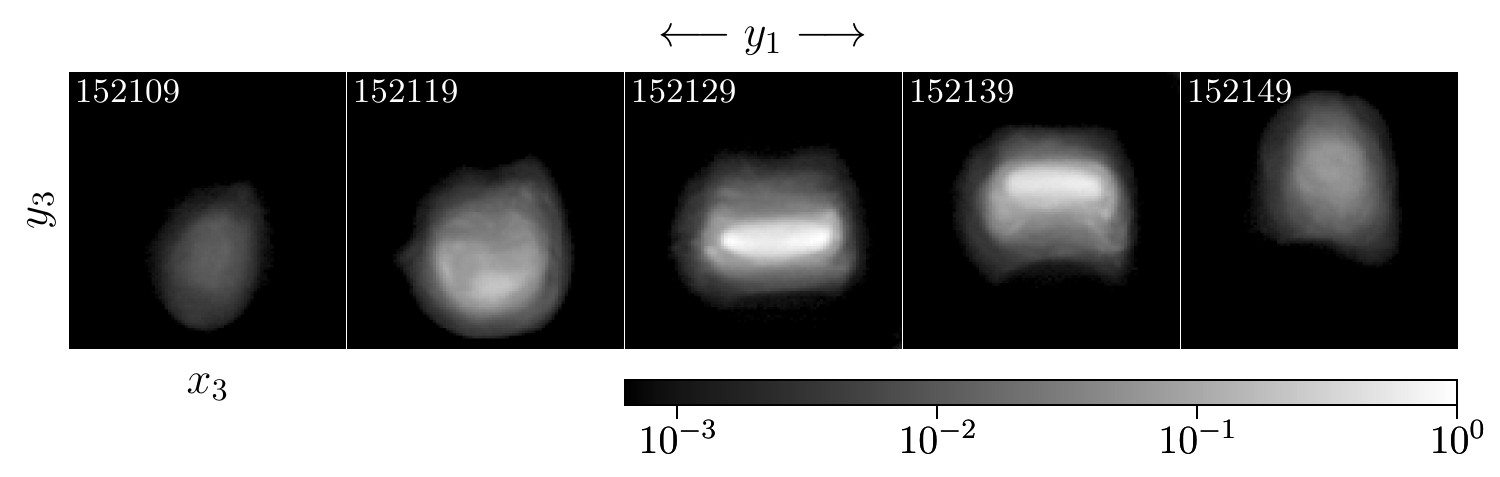}
    \caption{Processed camera images during one sweep. The vertical slits ($x_1$, $x_2$) are held fixed while the horizontal slit ($y_1$) is scanned.}
    \label{fig:scan_b}
\end{figure}

Images from the sweep containing the maximum camera integral are shown in Fig.~\ref{fig:scan_b}, which corresponds to one spike in the inset panel of Fig.~\ref{fig:scan_a}. All images were cropped, thresholded, and downscaled by a factor of three using local averaging. The resulting points and scalar values in five-dimensional slit-screen space were then linearly interpolated on a regular grid in five-dimensional phase space. After cropping, this procedure yielded a five-dimensional image of shape 69 $\times$ 88 $\times$ 69 $\times$ 65 $\times$ 55, with pixel dimensions 0.22 mm $\times$ 0.21 mrad $\times$ 0.37 mm $\times$ 0.20 mrad $\times$ 3.35 keV. This resolution approaches the limit dictated by the 0.2 mm slit widths. A discussion of the measurement uncertainty (which we judge to be relatively small) is included in Appendix~\ref{app-uncertainty}.

\section{Results}\label{sec:results}

\subsection{Revisiting the dependence of the energy distribution on the transverse coordinates}\label{sec:results-a}

Identifying and visualizing features in high-dimensional distributions is not straightforward \cite{Liu2017}. Although metrics are available to compare two distributions to each other \cite{Loudin2003, Modarres2020, Wong2022-symmetry, Wong2022-tomography, Mitchell2022}, it can be difficult to correlate these values with physical features. Visual inspection is a powerful tool but requires the distribution to be projected onto a one- or two-dimensional subspace.

\begin{figure}[]
    \centering
    \includegraphics[width=1.0\columnwidth]{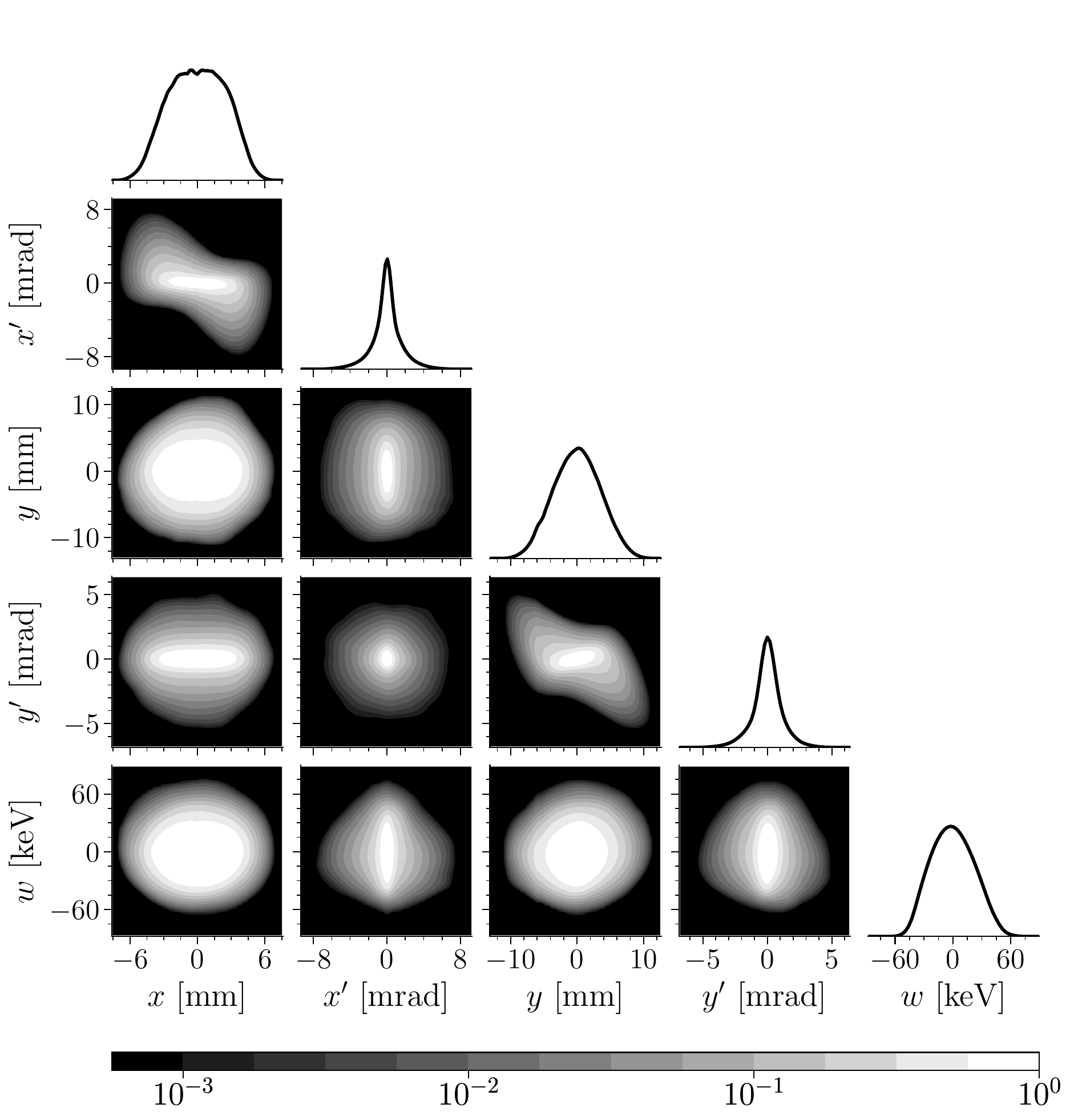}
    \caption{Corner plot of the measured five-dimensional phase space distribution. The one-dimensional projections are displayed on the diagonal subplots. Logarithmic contours of the two-dimensional projections are displayed on the off-diagonal subplots.}
    \label{fig:corner}
\end{figure}

The orthogonal one- and two-dimensional projections of the measured distribution are shown in a \textit{corner plot} in Fig.~\ref{fig:corner}.
No sharp features are visible, and all linear inter-plane correlations are negligible. One notable feature in the $x$-$x’$ and $y$-$y’$ projections is that the Twiss parameters in the core and tails/halo are dissimilar, which suggests that a matched core could lead to a mismatched halo. Some of these projections can be examined with a much larger dynamic range, as demonstrated in \cite{Aleksandrov2021}.

The projections in Fig.~\ref{fig:corner} represent averages over large regions of phase space and do not fully describe the distribution \footnote{It is helpful to observe the wealth of information contained in the two-dimensional projections in Fig.~\ref{fig:corner} relative to the one-dimensional projections. This suggests that the information lost during the transition from five/six dimensions to two dimensions could be significant.}. It is therefore critical to observe \textit{partial projections} \cite{Cathey2018, Ruisard2020}, where a partial projection is a projection of the distribution within some constrained region of phase space. When the region is small, the information loss is minimized, and a local description of the distribution follows; many such regions must be compared to build a global description. The selected region may generally be called a \textit{slice}. Slices are typically \textit{planar}; in an $n$-dimensional space, a planar slice selects an ($n - m$)-dimensional region defined by the intersection of $m$ orthogonal ($n - 1$)-dimensional planes. In practice, infinitely thin slices are not possible; for example, in the measurement described here, the slice width is limited by the physical slit widths. Thus, a planar slice is more practically defined as the intersection of orthogonal $n$-dimensional slabs.

\begin{figure}
    \centering
    \includegraphics[width=\columnwidth]{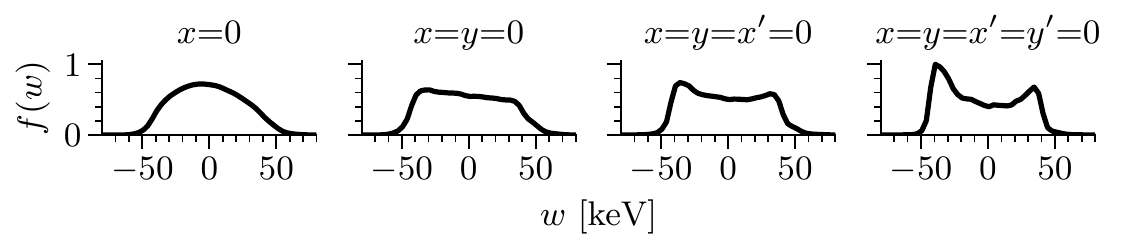}
    \caption{Energy distribution within planar slices in transverse phase space. Each slice is obtained by fixing the indices along the specified axes of the five-dimensional image. Each profile is normalized by area. (This figure mirrors Fig.~5 in \cite{Cathey2018}.)}
    \label{fig:hollow_energy_a}
\end{figure}

There is significant and largely unexplored freedom here, both in the slice construction and in the visualization of the resulting partial projections. We will revisit the previously observed longitudinal hollowing in the transverse core of the beam to accentuate this freedom. As mentioned in Section~\ref{sec:introduction}, this feature has thus far been examined by observing the energy distribution within a planar slice centered on the origin in transverse phase space, collapsing the slice dimensions one by one as in Fig.~\ref{fig:hollow_energy_a}.
We suggest two approaches to more comprehensively visualize this feature in five-dimensional phase space.\footnote{Each subplot in Fig.~\ref{fig:hollow_energy_a} represents a different subspace, ranging from two-dimensional to five-dimensional from left to right (neglecting the finite slice widths).}

\begin{figure*}[]
    \centering
    \includegraphics[width=\textwidth]{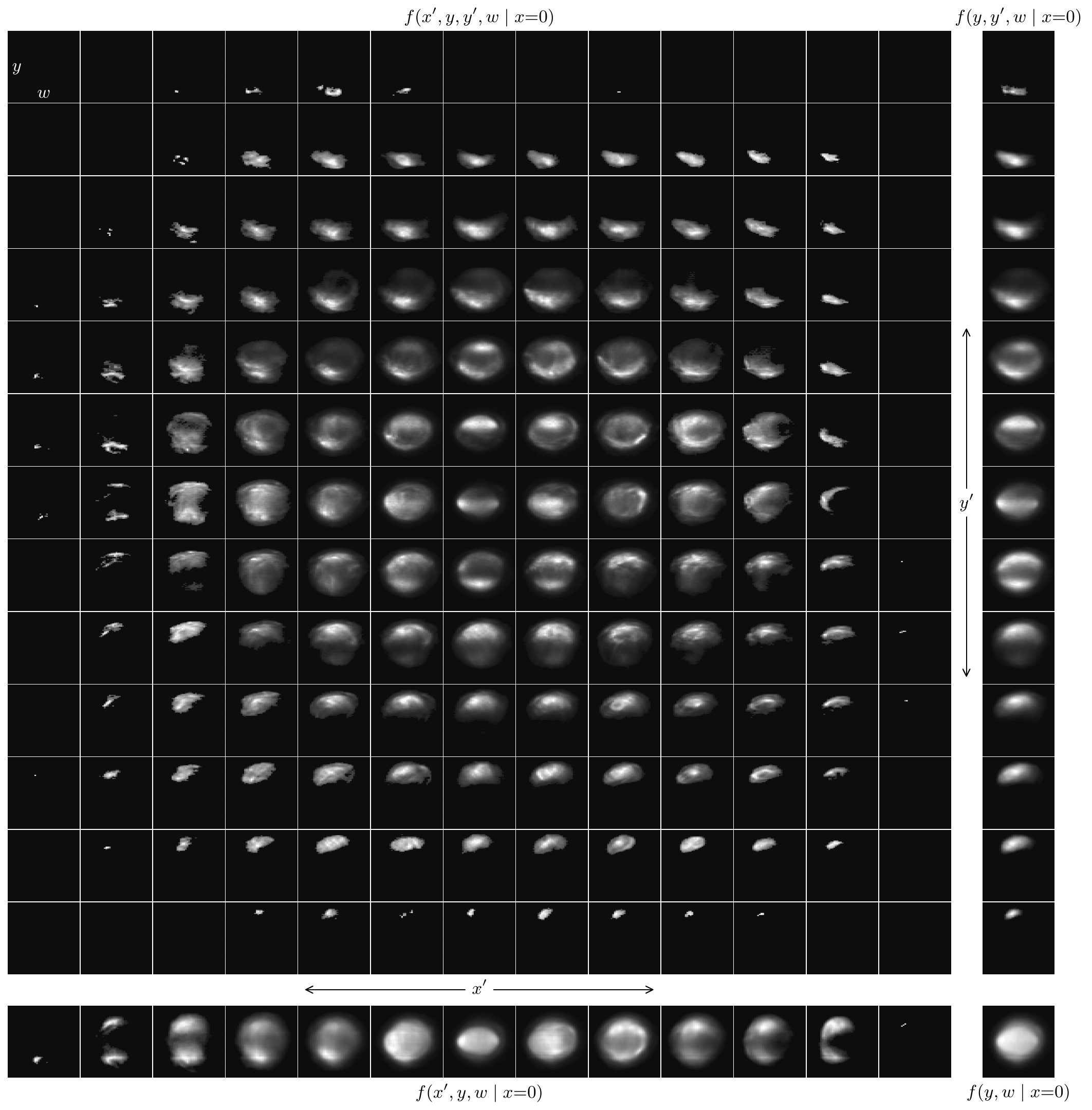}
    \caption{Dependence of the $y$-$w$ distribution on $x'$ (columns) and $y'$ (rows) near $x = 0$. Upper left: $f(x', y, y', w \mid x{=}0)$; upper right: $f(y, y', w \mid x{=}0)$; lower left: $f(x', y, w \mid x{=}0)$; lower right: $f(y, w \mid x{=}0)$. The color scale is linear and is not shared between frames. The axis limits are shared. 13/88 indices are selected along the $x'$ axis, and 13/65 indices are selected along the $y'$ axis. Each image is centered on the $y$-$w$ origin. The sliced dimension label $x'$ is located at $x' = 0$, with $x'$ increasing from left to right; the sliced dimension label $y'$ is located at $y' = 0$, with $y'$ increasing from bottom to top.}
    \label{fig:matrix_slice_xpyp_view_yw}
\end{figure*}

The first approach leverages the fact that an $n$-dimensional image is an ($n-2$)-dimensional array of two-dimensional images. When $n = 3$, the images can be arranged in a row. When $n = 4$, the images can be arranged in a grid \cite{Ruisard2020}. In Fig.~\ref{fig:matrix_slice_xpyp_view_yw}, we follow this approach to examine the slice $f(x', y, y', w \mid x{=}0)$.
In the main panel, the $y$-$w$ distribution is plotted as a function of $x'$ and $y'$. The bimodal energy distribution is visible near $x'=y'=0$ (seventh row/column) but quickly disappears as one moves away from the sharp peak in the $x’$-$y’$ distribution. The $y$-$w$ distribution in these low-density regions is somewhat complex and challenging to interpret.\footnote{Note that a linear correlation exists between $y$ and $y'$; this explains the shifting location of the first-order $y$ moment as $y'$ varies.} We also display the three-dimensional and two-dimensional marginal distributions on the bottom/right panels of the figure. These marginal distributions highlight the information lost by integrating over momentum space. The energy hollowing is still present in the marginal distributions, but is not as pronounced; this is consistent with Fig.~\ref{fig:hollow_energy_a}.

We stress that Fig.~\ref{fig:matrix_slice_xpyp_view_yw}, which we call a \textit{slice matrix plot}, still excludes a significant amount of information. First, only a fraction of the indices along the sliced dimensions are shown. Second, since the distribution is five-dimensional, one is tasked with observing a three-dimensional array of $y$-$w$ images; thus, one should vary the slice location along the fifth dimension ($x$, in this case). Third, a separate set of figures can be produced for each of the ten pair-wise relationships in the data set. These considerations can lead to a proliferation of figures, and the problem is worse in six dimensions. Nonetheless, the combination of several slice matrix plots for one or more carefully selected four-dimensional slices can be an effective tool to reveal the internal structure of a high-dimensional distribution.

\begin{figure}
    \centering
    \includegraphics[width=0.96\columnwidth]{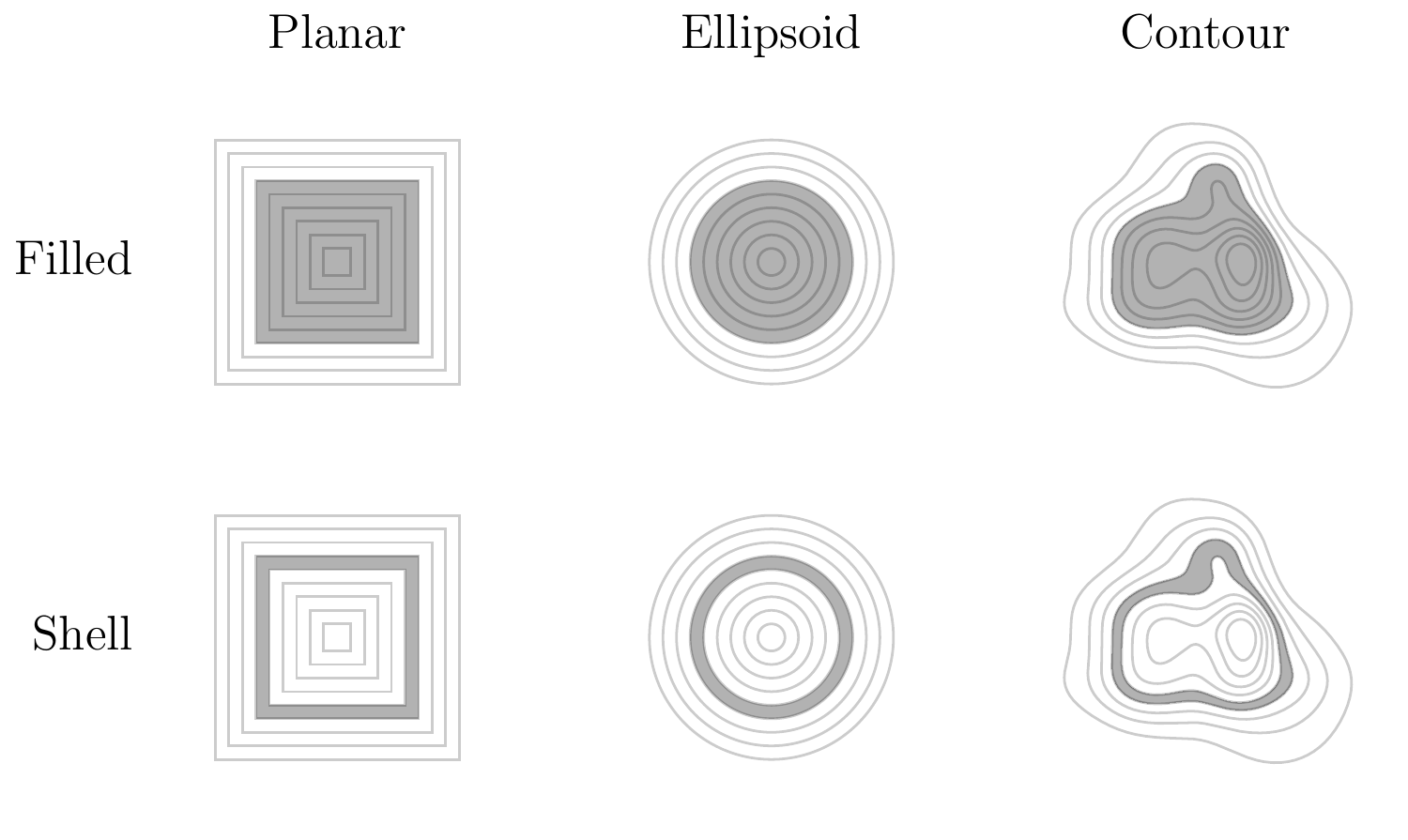}
    \caption{Several possible slice geometries. Each slice selects the shaded region of space.}
    \label{fig:slices}
\end{figure}

A second approach utilizes non-planar slices. Consider a slice of a distribution $f(x_1, x_2, \dots, x_n)$ defined by the intersection of $m$ perpendicular slabs, where $1 < m < n$ and slab $i\in[1, m]$ is defined by $|x_i| <= \Delta_i / 2$ for finite width $\Delta_i$. Let us refer to the $x_1$-$\dots$-$x_m$ plane as subspace \textit{A} and the $x_{m + 1}$-$\dots$-$x_n$ plane as subspace \textit{B}. In subspace $A$, the intersection defines an $m$-dimensional box of volume $V_A = \prod_{i}^{m}\Delta_i$. Instead of a box, one might consider an ellipsoid (perhaps defined by the covariance matrix of $f(x_1, \dots, x_m)$) or a more general boundary (perhaps defined by the density contours of $f(x_1, \dots, x_m)$). It is also possible to nest two such boundaries and select the region between them; we call this a \textit{shell slice}. Fig.~\ref{fig:slices} illustrates these options. In all cases, if the volume enclosed by the boundary goes to zero, we recover an ($n - m$)-dimensional planar slice. Note that for planar slices, it is generally advantageous to minimize $V_A$, but it may be advantageous to inflate the volume of non-planar slices.

\begin{figure}[]
    \centering
    \includegraphics[width=\columnwidth]{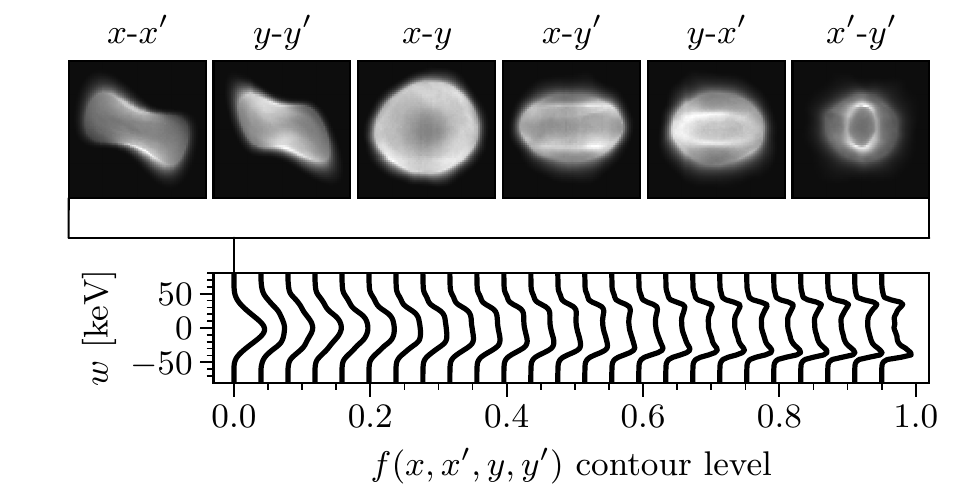}
    \caption{Bottom: energy distribution within contour shell slices in the $x$-$x'$-$y$-$y'$ plane. The slice at level $l$ selects the region $l \le f(x, x', y, y') \le l + 0.01$, with $f(x, x', y, y')$ normalized to the range [0, 1]. Top: two-dimensional transverse projections of the lowest density slice.}
    \label{fig:hollow_energy_b}
\end{figure}

Non-planar slices are well-suited to illuminate features in subspace $B$ that depend on the distance from the origin in subspace $A$. In particular, they are natural choices for demarcating the core and halo regions of the distribution \cite{Aleksandrov2016-IPAC}. There are many possibilities when applying these slices in six dimensions. (For example, one could select only those particles within the root-mean-square (RMS) ellipse in the two-dimensional longitudinal phase space and outside the $10^{-3}$ density contour in the four-dimensional transverse phase space, isolating the transverse halo in the longitudinal core.) In the case at hand, the energy distribution appears to have a radial dependence in transverse phase space, but it is clear that the transverse distribution does not have ellipsoidal symmetry. Therefore, we let the density contours of $f(x, x', y, y')$ define the slices. Each curve in Fig.~\ref{fig:hollow_energy_b} is the energy distribution within a shell defined by the region between two such nested contours.

Fig.~\ref{fig:hollow_energy_b} is compact but useful in describing the extent of the hollow energy core in the $x$-$x'$-$y$-$y'$ plane. The energy distribution transitions smoothly from unimodal to bimodal when moving from the low- to high-density contours. If the core is defined as the region in which $f(x, x', y, y') > 10^{-2}$, then the first slice selects the region outside the core, and subsequent slices select regions inside the core. (For reference, the 0.22 contour encloses one-fifth of the beam particles.) The two-dimensional projections of the lowest-density slice are shown in the top half of Fig.~\ref{fig:hollow_energy_b}. This slice essentially forms a contour-shaped shell around the beam core in the four-dimensional transverse phase space.

Since non-planar slices naturally identify the beam core and halo in high-dimensional phase space, they may be useful in future analyses, especially when the distribution lacks ellipsoidal symmetry. (One extension of the analysis shown here would be to vary the thickness of the shells --- averaging over a larger/smaller volume. Another extension would be to define the slices in a three-dimensional space; for example, viewing the $y$-$w$ distribution within contour slices in $x$-$x'$-$y'$ space.)

\subsection{Charge redistribution and core hollowing in the transverse plane}\label{sec:results-b}

\begin{figure}
    \centering
    \includegraphics[width=\columnwidth]{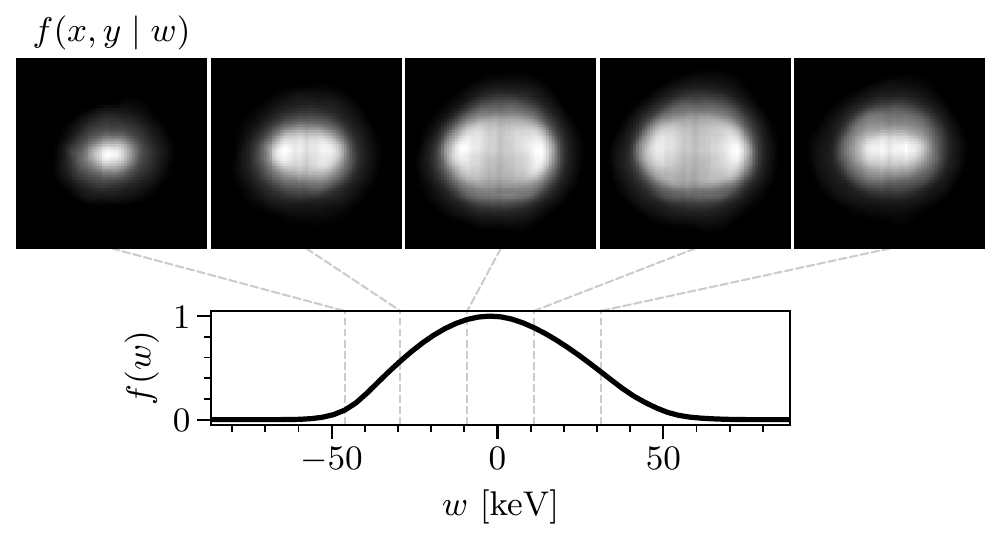}
    \caption{Dependence of the $x$-$y$ distribution on $w$. The color scale is linear and is not shared between subplots. Faint dashed lines indicate the location of each slice on the energy axis. The full energy projection $f(w)$ is shown on the bottom subplot.}
    \label{fig:wslices}
\end{figure}

We now explore the transverse phase space distribution and its dependence on the longitudinal parameters. The five-dimensional measurement has revealed an asymmetric, longitudinally dependent hollowing of the transverse charge distribution, shown in Fig.~\ref{fig:wslices}.

\begin{figure*}[]
    \centering
    \includegraphics[width=\textwidth]{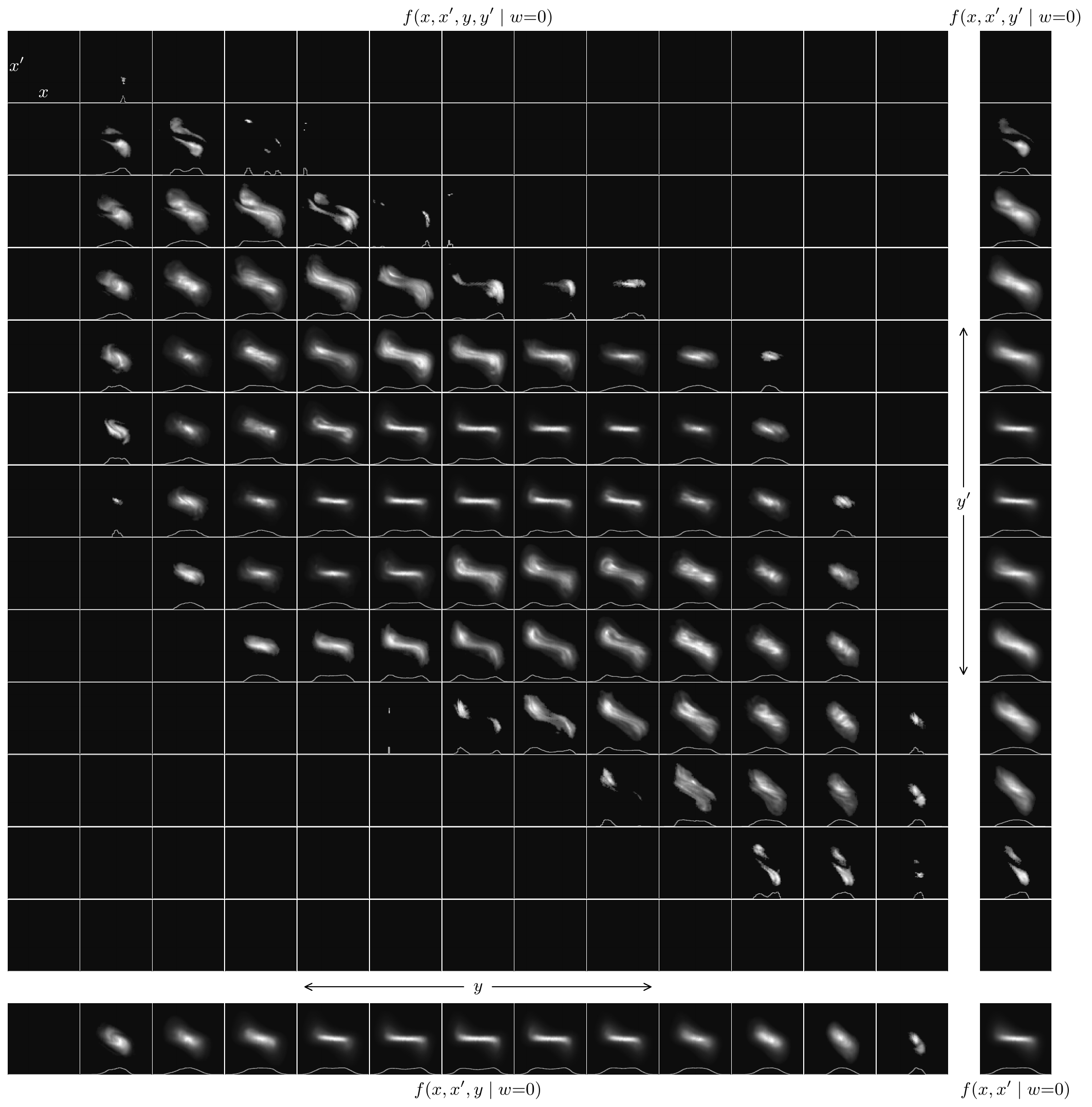}
    \caption{Dependence of the $x$-$x'$ distribution on $y$ (columns) and $y'$ (rows) near $w = 0$. Upper left: $f(x, x', y, y' \mid w{=}0)$; upper right: $f(x, x', y' \mid w{=}0)$; lower left: $f(x, x', y \mid w{=}0)$; lower right: $f(x, x' \mid w{=}0)$. The one-dimensional projection onto the $x$ axis is plotted as a white line. The color scale is linear and is not shared between frames. The axis limits are shared. 13/69 indices are selected along the $y$ axis, and 13/65 indices are selected along the $y'$ axis. Each image is centered on the $x$-$x'$ origin. The sliced dimension label $y$ is located at $y = 0$, with $y$ increasing from left to right; the sliced dimension label $y'$ is located at $y' = 0$, with $y'$ increasing from bottom to top.}
    \label{fig:matrix_slice_yyp_view_xxp}
\end{figure*}

\begin{figure*}[]
    \centering
    \includegraphics[width=\textwidth]{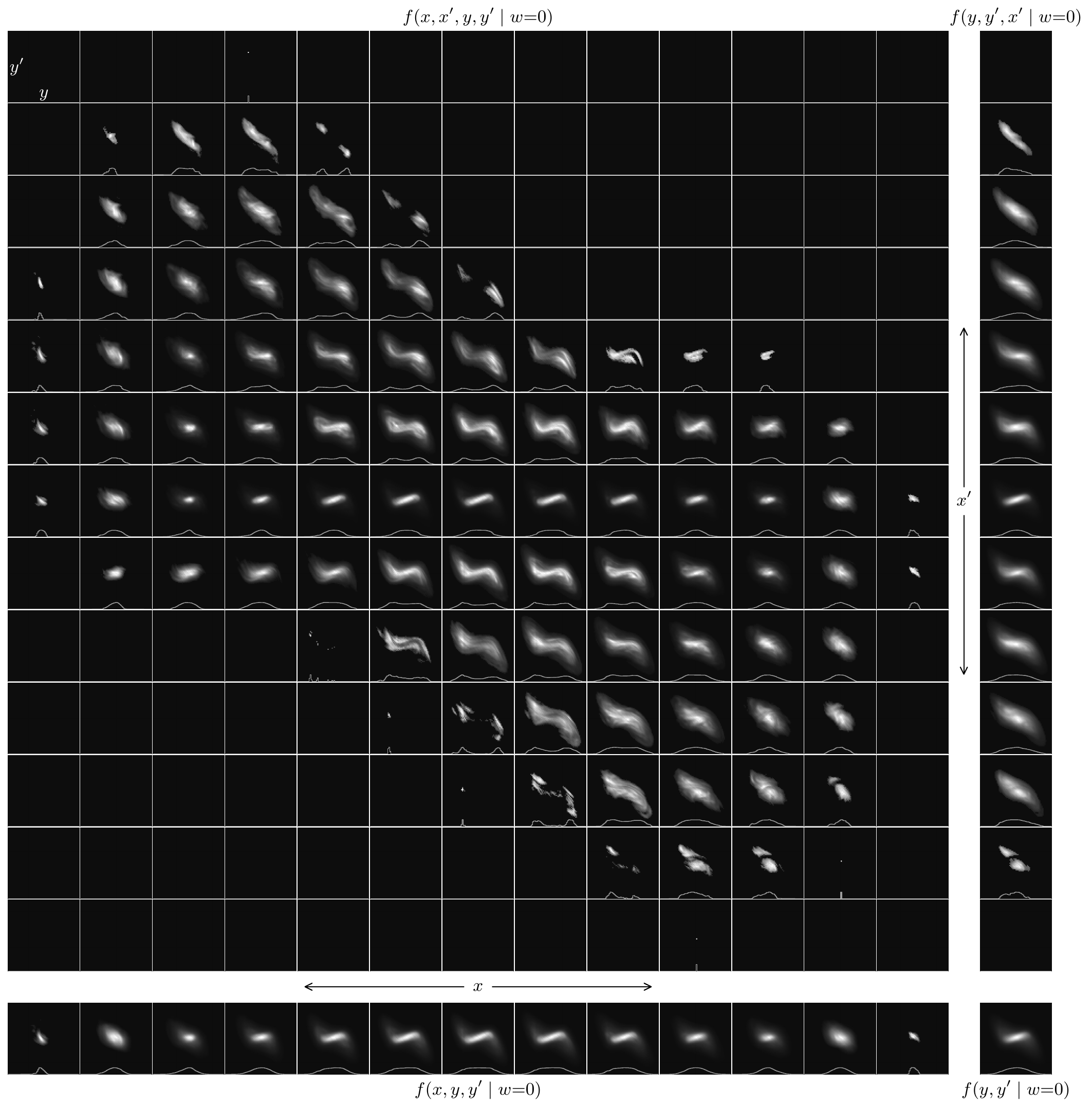}
    \caption{Dependence of the $y$-$y'$ distribution on $x$ (columns) and $x'$ (rows) near $w = 0$. Upper left: $f(x, x', y, y' \mid w{=}0)$; upper right: $f(x', y, y' \mid w{=}0)$; lower left: $f(x, y, y' \mid w{=}0)$; lower right: $f(y, y' \mid w{=}0)$. The one-dimensional projection onto the $x$ axis is plotted as a white line. The color scale is linear and is not shared between frames. The axis limits are shared. 13/69 indices are selected along the $x$ axis, and 13/88 indices are selected along the $x'$ axis. Each image is centered on the $y$-$y'$ origin. The sliced dimension label $x$ is located at $x = 0$, with $x$ increasing from left to right; the sliced dimension label $x'$ is located at $x' = 0$, with $x'$ increasing from bottom to top.}
    \label{fig:matrix_slice_xxp_view_yyp}
\end{figure*}

Some insight into the $x$-$y$ distribution can be gained by considering the four-dimensional transverse phase space. To this end, Fig.~\ref{fig:matrix_slice_yyp_view_xxp} shows the dependence of $x$-$x'$ on the vertical coordinates, and Fig.~\ref{fig:matrix_slice_xxp_view_yyp} shows the dependence of $y$-$y'$ on the horizontal coordinates, both within a central energy slice.
It is clear from these figures that a hollow $x$ or $y$ distribution is associated with the nonlinear tails of an ``s''-shaped $x$-$x'$ or $y$-$y'$ distribution. The asymmetric $x$-$y$ hollowing is explained as follows: after integration over $y'$ (bottom row in Fig.~\ref{fig:matrix_slice_yyp_view_xxp}), the $x$-$x'$ distribution near $y = 0$ is oriented such that the $x$ projection is bimodal; after integration over $x'$ (bottom row of Fig.~\ref{fig:matrix_slice_xxp_view_yyp}), the $y$-$y'$ distribution near $x = 0$ is oriented such that the $y$ projection is not bimodal.

The main panels of Fig.~\ref{fig:matrix_slice_yyp_view_xxp} and Fig.~\ref{fig:matrix_slice_xxp_view_yyp} indicate that there are inter-plane relationships in the transverse phase space distribution that are hidden by full projections. The orientation of the $x$-$x'$ distribution depends on the vertical phase space coordinates, and vice versa: the vertical distribution is diverging(converging) inside(outside) the $x$-$x'$ core. The shape of the $x$-$x'$ distribution depends on the vertical phase space coordinates, and vice versa: the ``s'' shape in one phase plane is most distinct near the origin in the other phase plane. 

One curious feature is the apparent ``splitting'' of phase space near the beam edge. This is visible in both Fig.~\ref{fig:matrix_slice_yyp_view_xxp} and Fig.~\ref{fig:matrix_slice_xxp_view_yyp} (for example, the frames at (row, column) = (5, 2), (3, 4) in Fig.~\ref{fig:matrix_slice_yyp_view_xxp}). This apparently exotic splitting is a straightforward consequence of using planar slices to examine a four-dimensional phase space distribution with nonlinear inter-plane correlations. It should also be noted that this is a minor feature of the distribution, accentuated only by the variable color scale per subplot: the peak density in frame (3, 4) is less than $1\%$ of the peak density across all frames.

We suggest that the transverse hollowing in the BTF is driven by nonlinear space charge forces in the MEBT, after the RFQ, and is independent of the longitudinal hollowing that develops in the RFQ. This suggestion is based on particle-in-cell simulations of the beam evolution, described below.

\begin{figure*}%
    \centering
    \subfloat[][PARMTEQ]{%
        \includegraphics[width=0.66\columnwidth]{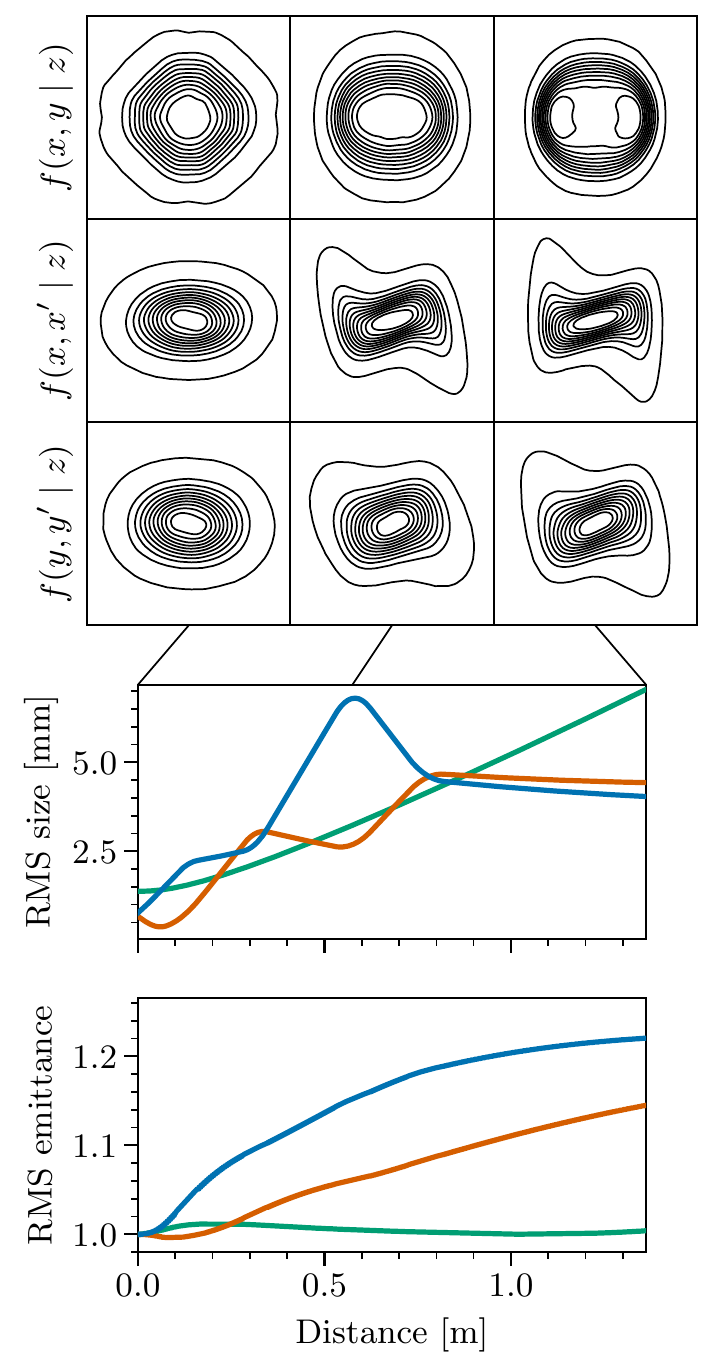}%
        \label{fig:sim_a}%
    }%
    \hfill
    \subfloat[][Gaussian]{%
        \includegraphics[width=0.66\columnwidth]{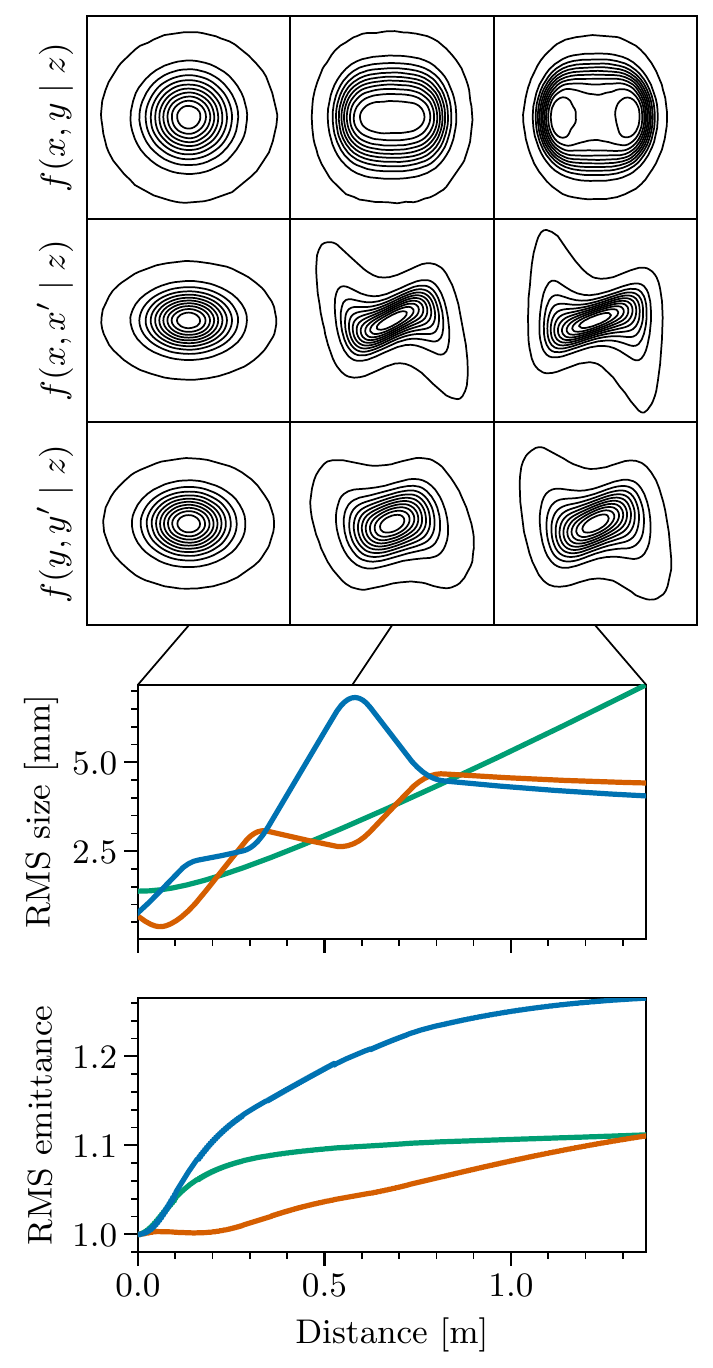}%
        \label{fig:sim_b}%
    }%
    \hfill
    \subfloat[][Waterbag]{%
        \includegraphics[width=0.66\columnwidth]{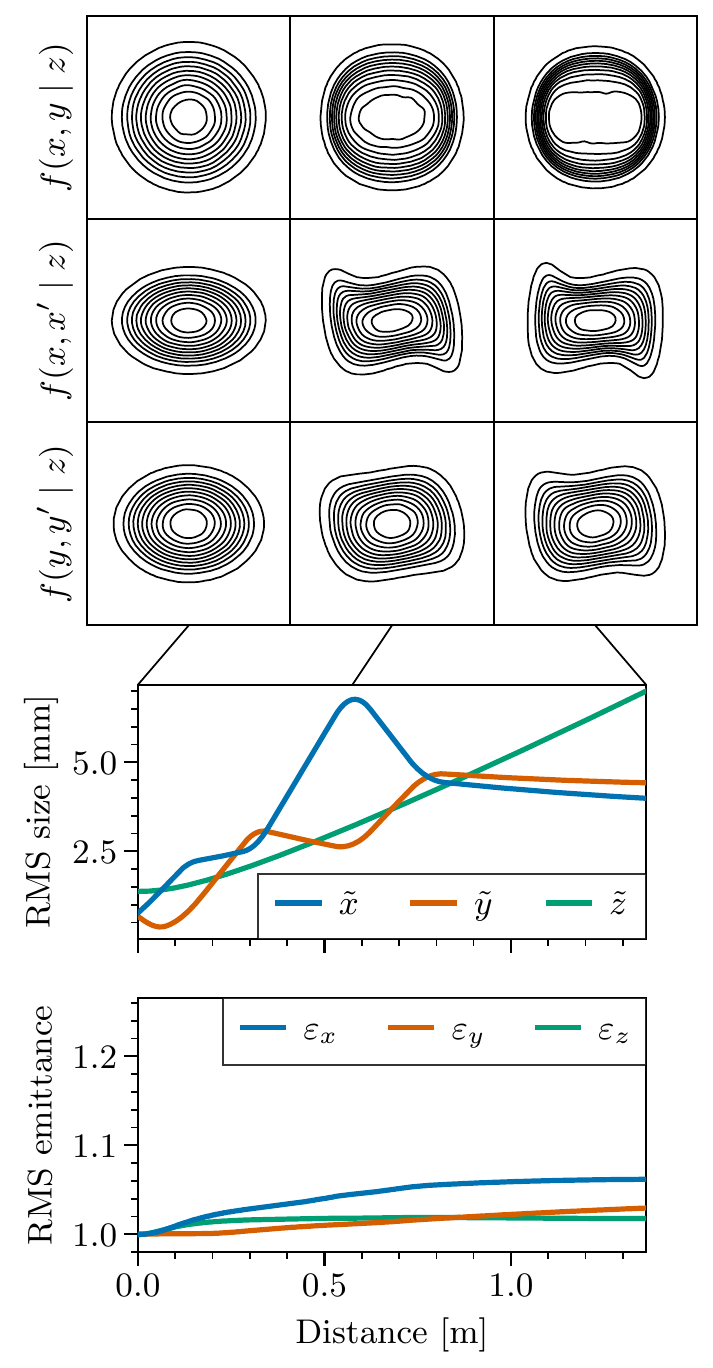}%
        \label{fig:sim_c}%
    }%
    \caption{Simulated transport of a 42 mA bunch in the BTF MEBT from the RFQ exit to the measurement plane (HZ04). (a) PARMTEQ-generated initial distribution. (b) Gaussian distribution, RMS-equivalent to (a) in the $x$-$x'$, $y$-$y'$, and $z$-$w$ planes. (c) Waterbag distribution, RMS-equivalent to (a) in the $x$-$x'$, $y$-$y'$, and $z$-$w$ planes. The top three rows show snapshots of $f(x, y \mid z\approx0)$ $f(x, x' \mid z\approx0)$, and $f(y, y' \mid z\approx0)$ at three locations in the lattice. Each distribution was normalized such that $\langle{xx}\rangle = \langle{yy}\rangle = 1$ and $\langle{xx'}\rangle = \langle{yy'}\rangle = 0$, where $\langle\dots\rangle$ represents the average over the distribution. Each set of contour lines was obtained by binning the coordinates on a $75 \times 75$ grid, then smoothing the resulting image using a Gaussian filter with $\sigma = 1.25$. Each set of contour lines range from 0.005-1.0 as a fraction of the peak density. The bottom two rows display the evolution of the RMS beam sizes ($\tilde{x} = \sqrt{\langle{xx}\rangle}$, $\tilde{y} = \sqrt{\langle{yy}\rangle}$, $\tilde{z} = \sqrt{\langle{zz}\rangle}$) and relative growth in RMS emittances ($\varepsilon_x = \sqrt{\langle{xx}\rangle\langle{x'x'}\rangle - \langle{xx'}\rangle^2}$, $\varepsilon_y = \sqrt{\langle{yy}\rangle\langle{y'y'}\rangle - \langle{yy'}\rangle^2}$, $\varepsilon_z = \sqrt{\langle{zz}\rangle\langle{ww}\rangle - \langle{zw}\rangle^2}$). Here we use the position $z$ instead of the phase $\phi$.}
    \label{fig:sim}
\end{figure*}

Our simulation procedure is described in detail in \cite{Ruisard2020}; we mention only the basic parameters here. The input bunch at the MEBT entrance was predicted using a PARMTEQ \cite{Crandall1988a} model of the RFQ. The input to the PARMTEQ simulation was based on two-dimensional phase space measurements in the LEBT at 50 mA beam current. The RFQ vane voltage was increased by 9\% over the design value of 83 kV based on preliminary results from x-ray spectrometry, which increased both transverse emittances by approximately 7\% at the RFQ exit. The predicted RFQ transmission was 84\%, resulting in a 42 mA beam current in the MEBT.\footnote{The transmission of the RFQ used in this study is lower than its design value due to gradual performance degradation during fifteen years of operation in the SNS. The exact reason for this degradation is unknown.} A PyORBIT \cite{Shishlo2015} model was used to propagate the bunch 1.3 meters from the RFQ exit to the first horizontal slit (HZ04), a distance including four quadrupole magnets for which a hard-edge model was used. Space charge kicks were applied every 2.5 millimeters using an FFT Poisson solver on a $64 \times 64 \times 64$ mesh with $8.6 \times 10^6$ macro-particles. Fig.~\ref{fig:sim_a} shows the simulated evolution, along with an RMS-equivalent Gaussian distribution in Fig.~\ref{fig:sim_b} and an RMS-equivalent Waterbag distribution in Fig.~\ref{fig:sim_c}.

A detailed study of the beam dynamics is beyond the scope of this paper; we briefly note the following conclusions drawn from the simulations.
\begin{enumerate}
    
    \item The transverse hollowing is qualitatively reproduced.
    
    \item The transverse hollowing develops in the MEBT regardless of the correlations that develop in the RFQ. This is supported by repeating the simulation after decorrelating the initial bunch by randomly permuting $x$-$x'$, $y$-$y'$, and $\phi$-$w$ coordinate pairs.
    
    \item The transverse hollowing is driven by nonlinear space charge forces. This is supported by comparing Fig.~\ref{fig:sim_b} and Fig.~\ref{fig:sim_c}: the hollowing (and resulting emittance growth) in the Waterbag distribution is reduced relative to the less-uniform Gaussian distribution. (The Waterbag distribution is approximately uniform, but not yet bimodal, at the end of the simulation.) Similar projected phase space densities have been observed in the simulated transport of an out-of-equilibrium four-dimensional Waterbag distribution in an alternate-gradient focusing channel; see Fig.~5 in \cite{Lund2009}. Of course, the details of the evolution are sensitive to the initial beam perveance, emittance, and the lattice focusing strength.
    
    \item The asymmetry in the $x$-$y$ hollowing is primarily due to the vertical beam waist in the early MEBT. The round initial beam, which is diverging horizontally and converging vertically, passes a vertical waist before the first quadrupole, then expands in both planes. The horizontal emittance grows most rapidly just after this waist, while the vertical emittance shrinks, presumably due to coupling between the planes. If the initial $x$ and $y$ beam divergences are exchanged ($x' \rightarrow -x'$, $y' \rightarrow -y'$), the hollowing is seen in $y$, not $x$, with an associated larger vertical emittance growth. The dependence on the exact pattern of alternate-gradient focusing is weak.
    
    \item The second-order moments disagree with the measurement --- for example, the RMS emittances differ by over 15\% --- even if the beam current is artificially decreased to the measured value of 25.5 mA. This is expected based on previous longitudinal benchmarks \cite{Ruisard2020}. A more detailed comparison with measurements is contained in \cite{Ruisard2022-NAPAC}.
    
\end{enumerate}

\begin{figure}
    \centering
    \includegraphics[width=\columnwidth]{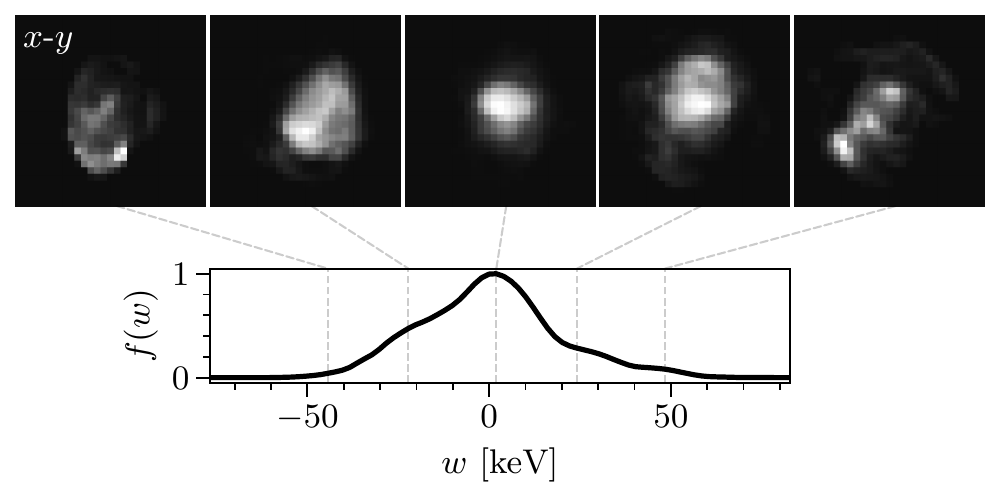}
    \caption{No transverse hollowing is apparent at the center of the energy distribution in the low-current (7 mA) measurement. This figure is equivalent to Fig. \ref{fig:wslices}, which shows the 26 mA case.}
    \label{fig:wslices_low_current}
\end{figure}

The simulations described above support the claim that the transverse hollowing is driven by nonlinear space charge forces in the MEBT. It is difficult to verify this claim experimentally without a current-attenuating grid immediately after the RFQ. Instead, we repeated a five-dimensional measurement at a lower beam current extracted from the ion source. This mirrors previous efforts to verify the space-charge-dependence of the longitudinal hollowing \cite{Cathey2018}. Fig.~\ref{fig:wslices_low_current} shows that no transverse hollowing occurs at this lower beam current.
Although the low-current five-dimensional distribution is not hollow, it is rich in structure, presumably due to the lack of smoothing by strong space charge. We leave the investigation of this low-current distribution as future work.

\section{Discussion}\label{sec:discussion}

In summary, we have used five-dimensional measurements to enhance our image of the initial phase space distribution in the SNS-BTF. We developed several high-dimensional visualization techniques and used them to re-examine the longitudinal hollowing in the transverse core. We also reported a transverse hollowing in the longitudinal core and explained its origin: simulations suggest that this feature is driven by nonlinear space charge forces in the MEBT, independent of the longitudinal hollowing that develops in the RFQ. We examined both features in considerable detail, leveraging the resolution and dynamic range of the five-dimensional measurement. Neither feature is visible in the two-dimensional projections of the distribution. Our data is further evidence that the three phase planes --- $x$-$x'$, $y$-$y'$, $\phi$-$w$ --- are (nonlinearly) correlated in real beams.

A longstanding goal in accelerator physics is to predict the beam evolution at the halo level, which we expect will hinge on (i) improving the accuracy of the accelerator lattice model and (ii) generating a more realistic initial bunch. Five-dimensional measurements at the end of the BTF beamline will serve as precise benchmarks and help address (i). To address (ii), direct six-dimensional phase space measurements are the gold standard. However, their demonstrated resolution and dynamic range are quite low.\footnote{Improved six-dimensional measurements may be available in the future: (i) we plan to increase the beam repetition rate from 5 Hz to 10 Hz; (ii) a smarter scan pattern could provide a factor of 5 to 10 speedup, as roughly 10\% of the points in a five-dimensional grid measure signal at the current dynamic range; (iii) a two-dimensional BSM would reduce the scan from five to four dimensions; (iv) the dynamic range can be increased by improving the light collection system \cite{Aleksandrov2021}. However, the level of detail is not likely to approach the high-resolution five-dimensional measurement described herein.} In the BTF, five-dimensional measurements may be able to serve as a proxy for six-dimensional measurements. For reasons described in Footnote~\ref{fn:1}, it is likely that a  reconstruction from \{$f(x, x', y, y', w)$, $f(\phi, w)$\} would be quite accurate. The reconstruction would ideally be treated using the principle of entropy maximization (MENT) \cite{Skilling1991}; a six-dimensional MENT solver could be adapted from one of several existing algorithms \cite{Skilling1984, Wong2022-tomography}. Alternative reconstruction approaches which incorporate low-resolution six-dimensional measurements may also be possible \cite{Dropulic2021}. In our case, it may suffice to sample from the five-dimensional distribution, then assume a linear relationship between $\phi$ and $w$, plus some phase width.

Our work may also be useful for high-dimensional phase space tomography --- the reconstruction of a four- or six-dimensional distribution from two-dimensional projections. There are various challenges in extending tomographic algorithms to six dimensions, mainly due to memory limitations, but also due to uncertainty in the set of transformations necessary to accurately reconstruct a high-dimensional distribution \cite{Hock2013, Wang2019, Wolski2020, Marchetti2021, Jaster-Merz2022, Wong2022-tomography, Wolski2022}. The accuracy of reconstruction algorithms has primarily been evaluated by comparing the two-dimensional projections of the reconstruction to the ground truth; it is an open question whether the high-dimensional features presented herein can be recovered. Direct measurements could serve as valuable benchmarks. Although the manipulations necessary for six-dimensional tomography are not possible in the BTF, our five-dimensional measurement data \cite{Hoover2023_Zenodo} could be used as a benchmark in a simulated reconstruction.

Finally, we note that the analysis and visualization techniques described here could be applied to fully correlated distributions generated by particle-in-cell simulations. For example, these techniques may be used to study the dynamics preceding the longitudinal hollowing in the RFQ or to track the evolution of high-dimensional, nonlinear inter-plane correlations after the measurement plane. It may also be beneficial to coupled visualization techniques with quantitative metrics such as Hilbert-Schmidt correlations \cite{Mitchell2022}.

\section{Acknowledgements}
The authors acknowledge the contribution of the SNS operators in enabling long (16+ hours) periods of continuous measurement. 
This material is based upon work supported by the U.S. Department of Energy, Office of Science, Office of High Energy Physics. This manuscript has been authored by UT Battelle, LLC under Contract No. DE-AC05-00OR22725 with the U.S. Department of Energy. This research used resources at the Spallation Neutron Source, a DOE Office of Science User Facility operated by the Oak Ridge National Laboratory. The United States Government retains and the publisher, by accepting the article for publication, acknowledges that the United States Government retains a non-exclusive, paid-up, irrevocable, world-wide license to publish or reproduce the published form of this manuscript, or allow others to do so, for United States Government purposes. The Department of Energy will provide public access to these results of federally sponsored research in accordance with the DOE Public Access Plan (http://energy.gov/downloads/doe-public-access-plan).

\appendix

\section{Transformation from slit-screen coordinates to phase space coordinates}\label{app-transform}

The following transformation from five-dimensional slit-screen coordinates to phase space coordinates is obtained by assuming linear optics in the measurement region \cite{Cathey2018-thesis}:
\begin{equation}\label{eq:transform}
    \begin{aligned}
    x &= x_1, \\
    y &= y_1, \\
    x' &= \frac{x_2 - x_1}{L_1}, \\
    y' &= \frac{y_3 - y_1}{L_1 + L_2 + \rho + L_3}, \\
    \delta &= \frac{1}{\rho + L_3} 
    \left(
        x_3 + \frac{L_3}{\rho} x -
        \left({
            \rho - \frac{(L_1 + L_2) L_3}{\rho}
        }\right) x'
    \right).
    \end{aligned}
\end{equation}
$L_1$ is the slit-slit spacing (HZ04-HZ06, VT04-VT06); $L_2$ is the slit-dipole drift length (VT06-DH1); $L_3$ is the dipole-screen drift length (DH1-VS06); $\rho$ is the dipole bend radius; $x_1$ is the position of the first vertical slit (VT04); $x_2$ is the position of the second vertical slit (VT06); $y_1$ is the position of the horizontal slit (HZ04); $y_3$ is the vertical position on the screen; $x_3$ is the horizontal position on the screen; $\delta = 1 + p/p_0$, where $p$ is the momentum and $p_0$ is the momentum of the synchronous particle. It is then straightforward to compute the energy deviation $w$ from $\delta$.

\section{Measurement uncertainty}\label{app-uncertainty}

Quantitative uncertainty analyses of previous BTF measurements are found in \cite{Cathey2018-thesis, Ruisard2020} and are generally applicable to the five-dimensional measurement described in this paper. For example, it is known that only slight systematic errors arise from the dipole strength calibration, image pixel size calibration, and lattice model geometry used in the conversion to phase space coordinates (Eq.~\eqref{eq:transform}) \cite{Ruisard2020}. Our present study differs only in the use of a view screen (instead of slits) to measure $y'$ and $w$.

Independent slit-based measurements of projections involving $y'$ or $w$ agree with screen-based measurements; for example, $f(y, y')$ can be measured using two horizontal slits and a Faraday cup. The screen offers a higher resolution in both coordinates. The $y'$ resolution limit is dominated by the width of the upstream horizontal slit, which contributes a point spread of 0.1 mrad. This is halved from the slit-slit geometry because the slit-screen distance is more than twice the slit-slit distance. The screen could support an energy resolution of 0.3 keV (given the field of view and raw image resolution), slightly less than the estimated 0.4 keV resolution limit dictated by the finite vertical slit widths \cite{Ruisard2020}.

Errors may also arise during interpolation. In this and all previous measurements, the data were interpolated one dimension at a time, taking advantage of a rectilinear scan pattern. As a consequence of the high number of steps per scanning actuator, the interpolation grid and the measurement grid nearly overlap. Thus, interpolation errors were minimized.

\begin{figure}[]
    \centering
    \includegraphics[width=0.77\columnwidth]{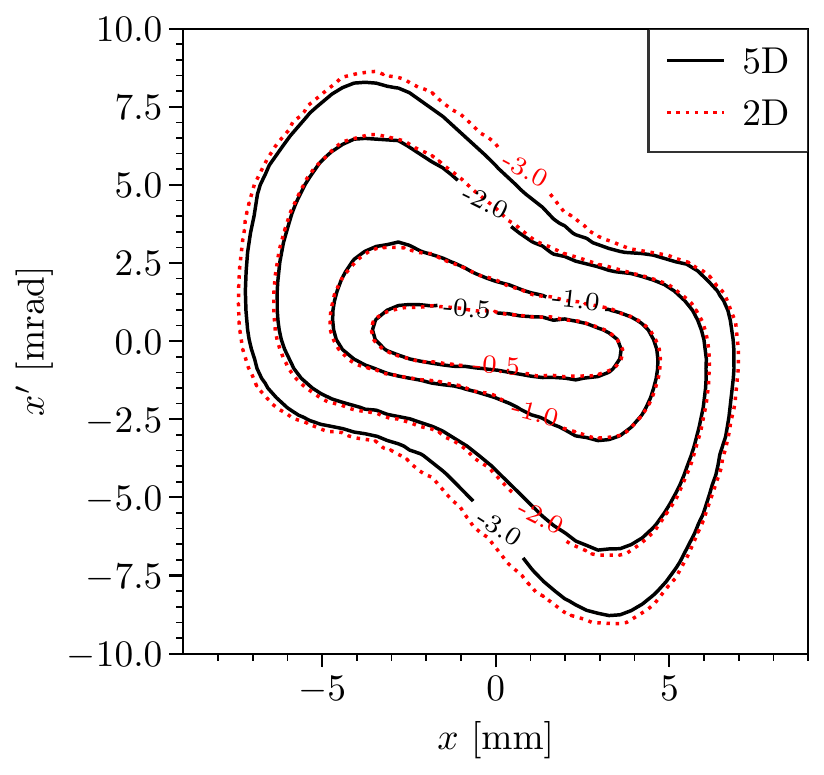}
    \caption{Logarithmic contours of $f(x, x')$ obtained from a seven-hour five-dimensional measurement (black) and seven-minute two-dimensional measurement (red) performed two weeks apart.}
    \label{fig:5D_vs_2D}
\end{figure}
\begin{figure}[]
    \centering
    \includegraphics[width=\columnwidth]{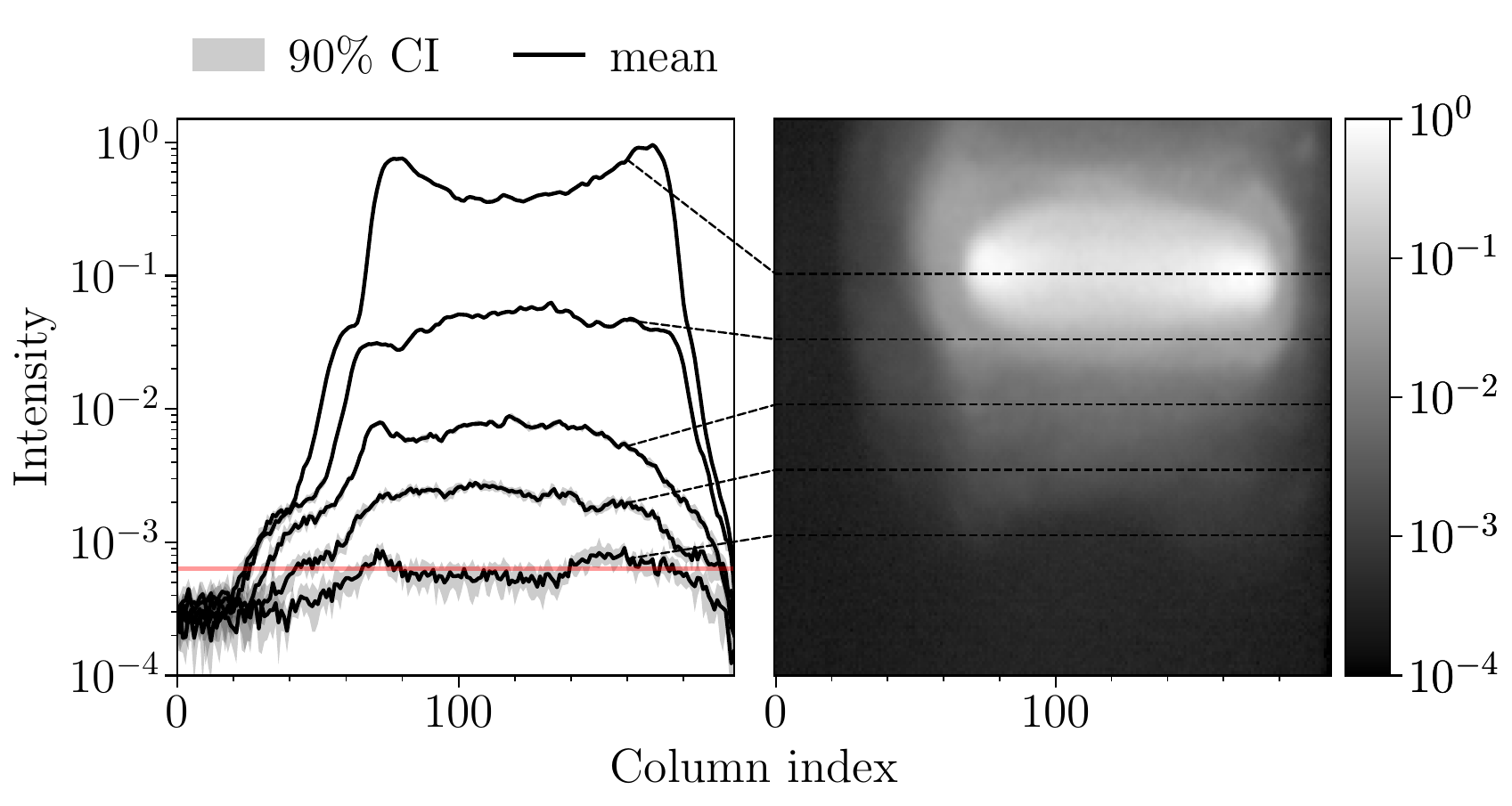}
    \caption{Pulse-to-pulse variation of the image on the screen. The image on the right represents $f(y', w \mid x{=}x'{=}y{=}0)$, averaged over ten pulses and downscaled by a factor of three. Several slices of the image are plotted on the left, showing the mean (black) and 90\% confidence interval (gray). The threshold used in the five-dimensional measurement is shown in red.}
    \label{fig:noise}
\end{figure}

Errors may also arise from pulse-to-pulse fluctuations of the phase space density. Averaging over multiple pulses is prohibitively slow in high-dimensional scans, so we instead use the ``flying slit'' scan described in Section~\ref{sec:introduction}. The following observations support the claim that the phase space density fluctuations were small enough to forgo averaging: (i) The beam current drift and jitter during the scan were negligible (Fig.~\ref{fig:scan_a}). (ii) A two-dimensional projection of the five-dimensional measurement agrees with a separate two-dimensional measurement down to three orders of magnitude (Fig.~\ref{fig:5D_vs_2D}). (iii) The pulse-to-pulse fluctuations of each each pixel in the five-dimensional image were estimated in a separate study. Three slits ($x$, $x'$, $y$) were inserted at the beam centroid before measuring ten pulses on the screen. Fig.~\ref{fig:noise} shows that fluctuations are only noticeable in very low-density slices, above the global threshold applied to the images in the five-dimensional measurement.

\bibliography{main}

\end{document}